\documentclass[10pt,english,a4paper]{article}
\usepackage{amssymb}
\usepackage{amsmath}
\usepackage{babel}
\usepackage{multicol}

\newcommand{\bee}{\begin{enumerate}}
\newcommand{\bei}{\begin{itemize}}
\newcommand{\beq}{\begin{eqnarray*}}
\newcommand{\beqn}{\begin{eqnarray}}
\newcommand{\ene}{\end{enumerate}}
\newcommand{\eit}{\end{itemize}}
\newcommand{\eeq}{\end{eqnarray*}}
\newcommand{\eeqn}{\end{eqnarray}}

\begin{document}
{\Large Is unbiasing estimators always justified ?}\\

{\footnotesize J.-M. L\'evy \\ 
\it Laboratoire de Physique Nucl\'eaire et de Hautes Energies,
CNRS - IN2P3 - Universit\'es Paris VI et Paris VII, Paris.   \it Email: jmlevy@in2p3.fr}\\

\begin{abstract}
It is argued that, contrary to common wisdom, unbiasedness is not always a well grounded
requirement. It is shown that in many cases, for a given unbiased estimator there is 
a simply derived biased estimator which gives results closer to the true value.
\end{abstract}
\begin{multicols}{2}
\section{Introduction and reminder}
For the purpose of devising the most efficient way of exploiting them, the results of physical experiments 
are generally regarded as realisations of random variables; statistical theory is then invoked to indicate 
efficient ways of using these sets of data for obtaining the values of physical parameters which are 
functionnally tied to the parameters of the probability distributions.  
In this framework, one calls {\bf estimators} certain random variables built on samples of potential observations 
which are used to evaluate part or all of the parameters of the underlying ('parent') probability distribution.
The sample average is amongst the simplest examples: if the expectation value $m$ of the parent distribution 
is unknown, the arithmetic mean $\bar{X}=\frac{1}{n}\sum_i X_i$ is the natural 
and, in many cases, the 'best' estimator that can be found to evaluate $m$ . The word
'best' has been quoted in the preceding sentence for reasons that will soon become 'clearer'. \cite{footnote1}\\

In many cases, the sample is (rightly) assumed to be made of independent observations
and -as we shall assume in the sequel- the underlying distribution is supposed to have moments up 
to second order: there exists an expectation value $E[X] = m$ and a variance
$V[X]=E[(X-m)^2] = \mu_2$ \cite{footnote2}
To define our notation, we call $A_n$ the estimator built on a sample of size $n$ and $a$ the 
parameter to be estimated, but we shall freely drop the subscript when it is irrelevant. We also assume that, 
as for $X$, the first two moments of $A$ exist.\\
\subsection{Estimator properties}
Since the {\bf estimation}, i.e. the value taken by the estimator after sampling, 
will be later used in place of the true value $a$, the one desirable property that $A$ 
should possess can vaguely be expressed by demanding it to take values as close as 
possible to $a$. How closeness is to be measured is the main issue in what follows.\\

Because it is easier to think in terms of fixed values rather than to keep in mind 
the full complexity of a probability distribution, one of the first ideas that comes to mind 
to satisfy this closeness requirement is to look for an estimator the expectation 
value of which is equal to the unknown parameter: $E[A_n] = a$. Such an estimator is
said to be {\bf unbiased}. When the bias (i.e. the difference $b_n = E[A_n] - a$) is not zero,
it often happens that it tends to zero when the sample size grows without limit, in which
case $A_n$ is said to be {\bf asymptotically unbiased} \\

{\bf At this point, it is important to stress 
that the only biases considered in the sequel are statistical biases, due to mathematical properties of the
estimators.} The measurements which are the source of the data can be 
affected by {\it \bf systematic} biases for instrumental reasons, a simple example of which being that
of a counter which misses part of the 'hits', thereby furnishing a systematically low count. {\bf We
assume that this kind of bias is being taken care of by appropriate means and we only address the question of the 
statistical biases in this article.}\\

Therefore, over and above unbiasedness, the first quality which is demanded for an estimator is {\bf consistency}:
a {\bf consistent estimator} must somehow approach the value to be estimated when
$n$ goes to infinity. The precise meaning of the word 'approach' in the previous sentence
can vary according to the kind of stochastic convergence which is adopted, but it is usually understood to refer to 
{\bf convergence in probability}, that is:  for {\bf any} given $\epsilon > 0$, the probability 
that $A_n$ deviates from $a$ by more than $\epsilon$ has zero limit when $n \rightarrow \infty$. More formally: 
$$ \forall \eta > 0 \;\; \exists N: \forall n > N \;\;P(|A_n-a| > \epsilon) < \eta$$
If $E[A_n]$ has a limit when $n \rightarrow \infty$, one does not see how that limit could differ from $a$ if the
previous requirement is fulfilled, but the author knows of no proof of this without additionnal assumptions.\\
The rationale for demanding consitency is fairly clear: it is 'obvious' (but can be false !) that averaging a
large number of measures of the same quantity will yield a better estimate of that quantity; on the other hand, 
accumulating data would be of no use if the estimation were not getting closer to the searched for value when the 
amount of data grows.

It is often written (see e.g.\cite{FRO}) that asymptotic properties such as consitency have nothing to say for finite 
sample sizes, contrary to unbiasedness which is a property defined for finite (read: 'realistic') sets of 
observations.
We think statements like this, supposedly based on good old common sense, are very deceiving; indeed, the observed 
average never equals its expectation value which is also, in a sense, an asymptotic property. All that can be said is that 
the average of an unlimited number of realisations  of $A$ converges to $a$ in some way. (The so-called {\bf law of 
large numbers}, more on this below)
But what is the relevance of all that for a 
single shot estimation built from a finite sample, especially if $A$ doesn't have a small dispersion ?
\cite{footnote3}
Concentration is therefore another important quality and one also demands the estimator to have a 'small' root mean 
squared that is, $\sqrt{V[A_n]}$ should not be larger than the error one is ready to tolerate on $a$ .\\
Building estimators with variances going to zero in the infinite sample limit is often possible in simple 
problems. Ideally, an estimator which is both asymptotically unbiased and of zero asymptotic variance is 
all that is required, would data be available in arbitrary large amounts: one easily shows that 
such an estimator is consistent by using Huygen's theorem: $$E[(A_n-a)^2] = V[A_n] + (E[A_n]-a)^2$$ and 
Chebyshev's inequality:$$P(|A_n-a| > \epsilon) < \frac{1}{ \epsilon^2}E[(A_n-a)^2]$$  
By the same token, one sees that if consistency is understood as convergence in {\bf quadratic mean}, it is 
completely equivalent to the conjunction of the two asymptotic requirements just stated.\\
 
All the insistence on expectation values and sample means comes from the above mentionned law of 
large numbers, of which there exist weak and strong varieties. For what concerns us, they both say that the 
arithmetic mean of $n$ equally distributed independent random variables converges in probability (weak law of l.n.) or 
almost surely (strong law of l.n.) towards their common expectation value when 
$n \rightarrow \infty$, as soon as this expectation value exists (analytically speaking); this explains why 
unbiasedness is expressed in terms of expectation values (but see note 3) and in simple cases, estimators 
are indeed averages of this kind for which the law applies.\\
However, although people can gather only finite samples, they tend to believe that their estimators will 
be 'better' if they are already unbiased for finite sample sizes; they often make big calculational
efforts to reach this aim - and spoil their estimators.  This is the belief and the practice that we challenge in the 
following.

\section{Why is unbiasing not necessarily a good idea.}
\subsection{Smaller variance or smaller bias ?}
There is a kind of trade-off between the two requirements of low bias and low variance in certain cases.
Let us assume that $A_n$ is multiplicatively biased, by this we mean that $ E[A_n] = f a$ where $f$ 
is some positive number $\neq 1$ which may be a function of $n$.\\
If $f$ is known, many practitioners of statistics will rather use $A'_n = A_n/f$. However, the variance of $A'_n$ is 
$V[A'_n] = V[A_n]/f^2$ and if $ f < 1$ one gets unbiasedness at the price of a larger dispersion 
{\bf and there is no reason to believe that $A'$ is better than $A$ only because its expectation value equals $a$}. 
Thinking so is somehow forgetting that a random variable is not its expectation value and unconsciously referring to 
the law of large numbers, which has, however, nothing to say about the relevance of an asymptotic property for a finite 
sample.
\subsection{What is closer ?}
Proximity will be dealt with in terms of distance, or difference. Definitions can vary and if the expected difference 
between 
$A'$ and $a$ is indeed zero by construction, the real life difference between the values taken by $A'$ and $a$ is 
never zero. Therefore it is more realistic to measure their distance by the mean absolute difference or the 
(root) mean square difference which is easier to handle, that is $D^2(A',a) \equiv E[(A'-a)^2]$ which is simply $V[A']$ 
for unbiased $A'$.\\
As for $A$,  Huygens' theorem says that: $D^2(A,a) = V[A] + (E[A]-a)^2$
The variance is the minimum of the mean square distance about a fixed point, and this minimum is reached for the 
fixed point taken at the expectation value. Therefore, if $A$ were additively biased, subtracting off the bias would
be the right thing to do. But this is not what we have in mind here.
\\ For $A'$, the squared distance to $a$ is $V[A'] = V[A]/f^2$\\
For $A$, it is $V[A] + a^2(1-f)^2$\\
The latter can be smaller than the former for $f < 1$ \cite{footnote4} 
and we shall base on this remark a general prescription for
improving estimators, but before so doing, let us examine a specific and well known example. 


\subsection{A simple example}
When it is required to estimate the variance of a distribution, the mean of which is
unknown, an 'obvious' estimator is the sample variance: \\$S^2 = \frac{1}{n}\sum_i(X_i-\bar{X})^2$ .
However, this $S^2$ is biased: $E[S^2] = \frac{n-1}{n} \mu_2$ which is precisely the kind of
situation that we are considering here. More often than not, people replace $S^2$ by $S'^2 = 
\frac{1}{n-1}\sum_i(X_i-\bar{X})^2 =\frac{n}{n-1}S^2$ which has obviously a larger dispersion.\\

To study the case further, let's make things simple and assume that the parent (sample) distribution is gaussian.
(The case of an arbitrary distribution is treated below)\\ 
$S^2$ is then the estimator of $\mu_2$ given by the maximum likelihood method when $m$ is unknown, but again, most
people shift to $S'^2$ because of the bias. However, it is particularly simple to show that one increases the dispersion 
of the estimator about $\mu_2$ by using this recipe.
\cite{footnote5}
Indeed, it is well known that $Q = \frac{nS^2}{\mu_2}$ is $\chi^2$-distributed with $n-1$ degrees of
freedom. Therefore $E[Q] = n-1$, $V[Q] = 2(n-1)$ and one has
$E[S^2] = \frac{n-1}{n} \mu_2$ as it has to be, $V[S^2] = \frac{2(n-1)}{n^2}\mu_2^2$ and
$V[S'^2]= \frac{2}{n-1}\mu_2^2$\\

But this entails that $$D^2(S^2,\mu_2) = V[S^2] + \mu_2^2(\frac{1}{n})^2 = \frac{2n-1}{n^2}\mu_2^2 < V[S'^2]$$
$S^2$ is therefore {\bf less dispersed about $\mu_2$} than $S'^2$ and it makes little sense to prefer the
latter on the grounds that it is unbiased. We can only disagree with, e.g. \cite{Eadie} who compare the 
bias with the loss in precision calculated as the difference between the standard deviations of the two estimates and 
settle the matter by claiming that 'for large $n$ this loss is very much smaller than the bias'. These are things
that cannot be compared. Of course, our mean square distance criterion makes use of expectation values just as 
the no bias criterion, {\it but a small $D$ is much more meaningfull than a zero expectation value; since all 
contributions are positive, they all add up in the calculation of $D^2$ and the true distance squared, in any given 
experiment, cannot be much larger than $D^2$ with any sizeable probability, whereas demanding no bias  
guarantees nothing of the kind since it can be
achieved by compensation of large opposite sign contributions.} \cite{footnote6} \\

One can derive limitations on the probability of 
an absolute difference from bounds on the variance or on the expected {\bf absolute} difference as 
examplified by Chebyshev's and Kolmogorov's inequalities. But nobody will ever succeed in deriving such a 
bound from a bound on the bias..to put it otherwise, the absolute value of the integral of a function has much less
to say about the size of that function than the integral of its absolute value.\\

Clearly, using $S^2$ will lead to average estimations below the true value of $\mu_2$ in the long run and the 
histogram built with many realisations of the Monte Carlo will not be 'centered' on the input value; many people 
would not like using $S^2$ precisely for that reason. We think the right answer is a flat:~~ 'So, what ?'
 The real question is: what are those estimates supposed to be used for ? If it is not to show colleagues how well 
you do in reconstructing the input parameters of your Monte-Carlo, then such things as those histograms should not
be considered as the primary criterion in assessing the quality of your estimators. People are taught and used to 
look at those features, but a minute of thought suffices to convince oneself that a centered histogram
proves very little. Control histograms can be plotted with unbiassed estimators to show that 'everything is 
understood', but that doesn't validate the estimators for whatever subsequent use is made of the estimates.\\
On the contrary, every student knows that, except for linear mappings, the expectation value of the transform is 
not the transform of the expectation value. Therefore, there is no real reason to insist on rigourously unbiased 
estimations. The perfectly legitimate requirement of being as close as possible to the true value is 
often contradictory with the 'no-bias' criterion.\\
To give yet another example: nobody would say that the mean distance to the origin in, e.g., a one dimensional, 
symmetrical random walk is zero on the grounds that the expectation value of the random walk is zero for even $n_{step}$. 
The root mean square is the universally accepted measure of distance, hence the $\sqrt{n_{step}}$ rule.


\section{If unbias doesn't help, what about..overbias ?}
\subsection{Optimal bias}
Having thus set foot in the marshes of heresy, going forward is the only logical attitude.
If $S^2$, above, is better than $S'^2$, what about $n/(n+k)S^2$ ?\\

Finding the optimum value of $k$ can be made by direct comparison: let $S''^2$ stand for the latter estimator 
Then $(V[S'^2]-D^2[S''^2,\mu_2])/\mu_2^2 = 
\frac{2}{n-1}(1-(\frac{n-1}{n+k})^2)-(\frac{k+1}{n+k})^2 = \frac{k+1}{(n-1)(n+k)^2}(3n+3k-nk-1)$\\
The largest difference obtains for $k=1$ and is equal to $\frac{4}{n^2-1}\mu_2^2$
The most 'concentrated' estimator about $\mu_2$ is therefore {\boldmath $S''^2 = 
\frac{1}{n+1}\sum_i(X_i-\bar{X})^2$}\\
This result is but a particular case of a more general formula that will be derived below.

\subsection{A word about error compensation}
Since the most 'concentrated' estimator of $\mu_2$ is $S''^2 = \frac{1}{n+1}\sum_i(X_i-\bar{X})^2$ and since this 
is probably not unknown, one might ask why people keep on using the unbiased $S'^2$ instead. 
Besides the already alluded to histograms, the unconscious idea underlying the use of unbiased estimates is
probably that fluctuations above and below the 'true value' (which is the expectation value of $S'^2$ in this case) 
should more 
or less compensate. We have already remarked that such a motivation is poorly grounded for a one-time estimation.
But for the sake of the argument, let us take the idea seriously. The best estimator in that case should be such that
its probability to be above the 'true value' is equal to its probability to be below this value. In other words,  
for our example, $\mu_2$ should be the {\bf median} of the distribution of the estimator rather than its mean. Let's 
therefore define an 'ideal' $S^2_{id}$ proportionnal to $S'^2$ such that $\mu_2$  be the median of its distribution.
Let $S^2_{id} = \alpha S'^2$ Then $\frac{(n-1)}{\alpha \mu_2} S^2_{id}$ is $\chi^2$ distributed with
$n-1$ d.d.f. and the condition we impose calls for finding the median ${\cal M}_{n-1}$ of the $\chi^2$ distribution. 
Numerical evaluation up to $n = 400$ shows that the median of 
$\chi^2_n$ is always between $n$ and $n-1$, slowly decreasing and seemingly converging towards $n-2/3$ but this 
value is, of course, only a guess. This means that $\alpha = \frac{n-1}{{\cal M}_{n-1}} > 1$ and therefore that 
$S^2_{id}$ is not below but above $S'^2$ contrary to the conclusion to which we were led by our distance argument. 
One has $S^2_{id} = \frac{1}{n-r} \sum_i(X_i-\bar{X})^2$ with $r \approx 5/3$ and the (mean squared) distance between 
$S^2_{id}$ and $\mu_2$ is $V[S^2_{id}] + (\frac{r-1}{n-r})^2\mu_2$, larger than everything found so far.\\

Facing this distressing result, one might think of a last way out for the case at hand: in the same spirit as our 
tentative use of the median and in line with the philosophy of the maximum likelihood method, one could assume that 
the best estimate of $\mu_2$ is that which renders the value found for $S'^2$ most likely. Contrary to the median case,  
it is quite easy to find by derivation that the most likely value (the so-called '{\bf mode}') of a $\chi^2_n$ 
distribution is $n-2$. Therefore the maximum likelihood estimator of $\mu_2$ 
{\bf in this sense} should be taken as $\frac{n-1}{n-3}S'^2 = \frac{1}{n-3} \sum_i(X_i-\bar{X})^2$, still  
farther away from $S^2$ than the preceding estimate (recall that $S^2$ {\bf is} the maximum likelihood estimate  
for a gaussian sample with unknown mean). The maximum likelihood method seems therefore to suffer of some 
kind of schizophreny: the maximum likelihood estimator of $\mu_2$ based on the full sample distribution, that is 
$S^2$, is not the same as the maximum likelihood estimator of the {\bf same } parameter based on the distribution of 
this {\bf same} $S^2$, which is $\frac{n}{n-3}S^2$ ..\\ 
\section{A general prescription}
\subsection{Improving an unbiased estimator}
The lesson of the latter section is that 'compensation' arguments lead only to contradiction. 
Even the time honored maximum likelihood method is shown to be self-inconsistent. Aware of this fact, some people use 
M.L. only as a starting point to find some estimator which they further 'improve'. But as already remarked, the supposed 
improvement can spoil the result and this is particularly clear on the example that we have used. \\

On the other hand, the minimum squared distance criterion is certainly better grounded than the no-bias prescription 
for reasons which have already been explained. One then might ask for a general rule based on it. \\
Using the notations of paragraph 2.2, the squared distance of $A$ to $a$ can be written:
$$D^2(A,a) = V[A] + (E[A]-a)^2 = f^2 V[A'] + a^2(f-1)^2$$
Therefore, having found an unbiased estimator $A'$ one can try to derive a smaller distance estimator by minimizing 
the above expression w.r.t. $f$. Zeroing the derivative yields the condition: $$f = f_m \equiv 
\frac{a^2}{E[A'^2]}$$ where use
has been made of $V[A'] = E[A'^2] - a^2$. Since $a^2 = E[A']^2 <  E[A'^2]$, $f_m < 1$ as expected.
Starting from any unbiased $A'$, it is always possible to build, in principle, an improved estimator:
$$A_{better} = \frac{a^2}{E[A'^2]} A'$$  which will be closer to the unknown parameter than $A'$.
It is, of course, biased, but its squared distance to $a$ is easily seen to be reduced by a factor of $f_m$\\ 
Note that, at least for 'mean square' consistency, $f_m \rightarrow 1$ when $n \rightarrow \infty$ because 
convergence in the mean square entails convergence of the first two moments
of the distribution towards those of a constant; in particular, $E[{A'}^2_n] \rightarrow E[a^2] = a^2$ so that
as soon as they are built from consistent estimators, our biased estimators are themselves consistent.\\
The expression found for $A_{better}$ seems to depend on the value to be estimated. However, alhough there exist 
indeed cases in which it is of no use, we'll see presently that there are some important 
simple problems where it is perfectly usable.\\
Moreover, even if the exact value is not known, any non trivial upper bound on $f_m$ (that is, smaller than 1) will 
yield some improvement if used in place of $f_m$ to bias $A'$.\\
It is important to observe here that the bias so introduced always tends to zero when $n \rightarrow \infty$
as soon as the (unbiased) estimator variance goes to zero. Indeed, $f_m =  \frac{a^2}{V[A']+a^2} \rightarrow 1$
in this case. 

\subsection{Examples}
\begin{itemize}
\item Estimation of the variance of a gaussian distribution with unknown expectation value.\\
This is the already treated example. Here $a = \mu_2$ and $E[A'^2] = V[S'^2] + \mu_2^2 = \frac{2}{n-1}\mu_2^2 + 
\mu_2^2$ therefore $f = \frac{n-1}{n+1}$ and $S^2_{better} = \frac{1}{n+1}\sum_i(X_i-\bar{X})^2$ as already found.
\item  Estimation of the variance of a gaussian distribution with known expectation value.\\
The unbiassed estimator is here \\$S_0^2 = \frac{1}{n}\sum_i(X_i-m)^2$ \\with $V[nS^2/\mu_2] = 2n$ hence $V[S_0^2] = 
\frac{2\mu_2^2}{n}$ \\
Therefore $f = \frac{n}{n+2}$ and \\$S^2_{better} = \frac{1}{n+2} \sum_i(X_i-m)^2$  
\item A variation on the first example can be found in the problem of the linear least square fit with gaussian
errors. When the overall scale $\sigma^2$ of the covariance matrix $V = \sigma^2 W$ of the observations is unknown, 
finding the parameter estimators is still possible ($W$ is assumed to be known), but not so for their covariance 
matrix or for the variance of a prediction. One can then estimate $\sigma^2$ using the fact that the residual quadratic 
form $Q_{min}$ is $\chi^2$-distributed with $n-k$ degrees of freedom, with $n$ the number of points and $k$ the number of 
estimated parameters (see e.g. \cite{FRO2} ). One has $Q_{min} = \frac{^t\epsilon W^{-1}\epsilon}{\sigma^2}$ 
with $\epsilon$ the vector of residuals, and an 
unbiased estimator of $\sigma^2$ is therefore $\hat{\sigma^2} = \frac{^t\epsilon W^{-1}\epsilon}{n-k}$. \\According 
to our recipe, $\hat{\sigma^2_{better}} = \frac{^t\epsilon W^{-1}\epsilon}{n-k+2}$
\item Estimation of the variance of an arbitrary distribution with known  expectation value.\\
With $S_0$ as here above for the unbiased estimator one finds: $E[(S_0^2)^2] = \frac{1}{n^2}E[\sum_i(X_i-m)^4
+2\sum_{i < j}(X_i-m)^2(X_j-m)^2] = \frac{1}{n}\mu_4+\frac{n-1}{n}\mu_2^2$ \\The improved estimator is therefore:
\\$S_{better}^2 = \frac{n}{\gamma + n -1}S_0^2$ with $\gamma$ the ratio  
$\frac{\mu_4}{\mu^2_2}$ ($\gamma$ equals $3$ for a gaussian distribution, which checks our preceding result).
\item Estimation of the variance of an arbitrary distribution with unknown  expectation value.\\
This calls for the more tedious calculation of the second moment of $S'^2$ defined above.
One finds $E[(S'^2)^2] = \frac{\mu_4}{n} +\frac{n^2-2n+3}{n(n-1)}\mu_2^2$ and the improved estimator can
be written: $$S'^2_{better} = \frac{n}{\gamma+n-1+\frac{2}{n-1}}S'^2$$ Note that by Schwartz's inequality,
$\gamma > 1$ in accordance with our calculations for the last two items. The second result yields a marginal 
improvement even in the absence of a better knowledge of $\gamma$ than this trivial bound.
    
\item Estimation of the parameter of an exponential distribution.\\
The density is $\frac{1}{\tau}e^{-\frac{t}{\tau}} \;\;{\rm for\;\;} t > 0\;$ and the unbiased estimator of $\tau$ 
is 
$\hat{\tau}= \frac{1}{n}\sum_i T_i$
with variance $\frac{\tau}{n}$ One finds $\hat{\tau}_{better} = \frac{1}{n+1}\sum_i T_i$
\item If, in the preceding exemple, one prefers to estimate the rate $\lambda = \tau^{-1}$, the M.L. estimator is 
$\hat{\lambda}= \frac{n}{\sum_i T_i}$ The moments of $\hat{\lambda}$ are easily computed by observing that
$\frac{n}{\hat{\lambda}}$ is distributed according to a $\gamma(n,\lambda)\;$ law and by using the 
normalisation integral: \\$\Gamma(k) = \int_0^{\infty} x^k\lambda^{k-1}e^{-\lambda x} 
dx$. $\hat{\lambda}$ is biased but $\frac{n-1}{n}\hat{\lambda}$ is not and one finds  
that the improved estimator is $\hat{\lambda}_{better} = \frac{n-2}{n}\hat{\lambda}$
\item For an unbiased estimator which reaches the minimum variance bound (Cramer-Rao inequality), the factor $f_m$
reads $\frac{a^2}{a^2 + 1/I_n(a)} = \frac{a^2 I_n}{1+a^2 I_n}$ where $I_n$ is the amount of 
information on $a$ brought by the $n$-sample, viz: $I_n(a) = E[(\frac{\partial Log {\cal L}}{\partial a})^2]$
with $\cal L$ the likelihood of the sample. The lifetime estimator above is a case of that kind. 
\item For a last example, let us consider the maximum likelihood estimator of the parameter $\theta$ of a
uniform distribution on $[0,\theta]$. This is $\hat{\theta} = sup_i{X_i}$ were the $\{X_i\}$ stands for the sample.
This estimator is easily shown to be biased: $E[\hat{\theta}] = \frac{n}{n+1}\theta$ The unbiased estimator is
therefore $\frac{n+1}{n}sup_i{X_i}$ from which one easily finds the improved $\theta_{better} = 
\frac{n+2}{n+1}sup_i{X_i}$
\end{itemize}
\section{Summary and conclusion}

It has been argued that the requirement of unbiasedness at the price of a larger mean square distance to the
estimated parameter is not well grounded. Mean absolute differences or mean squared differences are clearly more 
meaningfull than the average of signed differences which can hide large fluctuation through compensation. 
It has been observed that the demand of a 'centered' histogram is a matter of habit, but has no meaning as to the
optimality of the estimates for other purposes than checking calculations. Requiring histograms to be 'centered' in 
reference to the median would not be less legitimate. \\    
However, with the help of a definite example, it has been shown that attempts to use some kind of fantasmatic 'error 
compensation' through the use of mean, median or mode leads to contradictions and to the use of estimators with ever 
wider distributions.\\ On the other hand, minimizing the mean square distance gives a general prescription to improve
on an unbiased estimator by biasing it to a slightly lower expectation value. 
Even if the formula for the bias factor thus obtained is 
not always applicable because of the unknown quantities that it involves, it has been shown that it yields perfectly 
definite and usable results in some important cases. Any non trivial upper bound on this bias factor yields some
improvement.\\

In conclusion, unbiased estimators are certainly usefull for constructing control histograms, but should not be
automatically taken at face value when the problem is that of using the estimates for further calculations.

\begin{footnotesize}

\end{footnotesize}

\end{multicols}


\begin{thebibliography}{2} 
\bibitem{footnote1} ..and the word 'clearer' has been quoted for reasons which will pretty soon become obvious (no 
quotes)
\bibitem{footnote2} We use $\mu_2$ rather than the more traditional $\sigma^2$
here because we shall be led to use higher moments and a uniform notation is desirable.
The centered moment of order $k$ will be called $\mu_k$.
\bibitem{FRO} Frodesen, Skjeggestad, T\o fte {\it Probability and statistics in particle Physics}
Universitetsforlaget, Bergen, Oslo Troms\o, p.182.
\bibitem{FRO2} Ref. [3] p. 284
\bibitem{footnote3} Moreover, the standard definition of 'unbiasedness' is quite arbitrary. There is nothing sacred 
about expectation value and one could as well demand that the median 
${\cal M}[A_n]$ of the distribution of $A_n$ (defined by $P(A_n \leq {\cal M}[A_n]) = P(A_n \geq {\cal M}[A_n])= 
1/2$) be equal to $a$. 
After all, the median of the distribution of the sample median is equal to the median of the parent distribution, 
call it $\mu$, and the sample median is a consistent estimator of $\mu$ for distributions which do not even 
possess an expectation value (e.g. the lorentzian distribution)
\bibitem{footnote4}On the contrary $f > 1$ would entail a larger dispersion of $A$ around $a$ and is therefore uninteresting. 
\bibitem{footnote5} Quite apart from
these considerations, the case for $S^2$ is sometimes pleaded on the grounds that it is the maximum likelihood 
estimator among all those samples which have averages equal to the value of $\bar{X}$ which has been found. See 
[7]
\bibitem{K} M.G. Kendall {\it The Advanced Theory of Statistics}, 3d ed. vol II pp 34 ss.
\bibitem{Eadie} W.T. Eadie {\it et al} in {\it Statistical Methods in Experimental Physics} North Holland (1971)
p.181 
\bibitem{footnote6} Think of an unbiased estimator with a 
U-shaped probability distribution function ! This, by the way,
is what vindicates the demand for consistency.

\end{thebibliography}
\end{document}